\documentclass[runningheads,bookmarksdepth=2]{llncs}

\usepackage[T1]{fontenc}

\usepackage{mdframed}

\usepackage{graphicx}

\usepackage[svgnames,dvipsnames]{xcolor}

\usepackage{enumitem}

\usepackage{xspace}

\usepackage{tikz}
\usetikzlibrary{arrows.meta}

\usepackage{hyperref}
\hypersetup{
  bookmarksopen=true,    
  pdftitle={A Metrics-Oriented Architectural Model to Characterize Complexity on\break Machine-Learning Enabled Systems},
  pdfauthor={Renato Cordeiro Ferreira},
  pdfcreator={pdflatex}, 
  pdfnewwindow=true,     
  colorlinks=true,       
  linkcolor=blue,        
  citecolor=DarkGreen,   
  filecolor=green,       
  urlcolor=DarkRed       
}

\usepackage[nameinlink,noabbrev]{cleveref}

\newcommand{\etal}{\textit{et al.}\xspace}

\newcommand{\OG}{\textsc{Ocean Guard}\xspace}
\newcommand{\SPIRA}{\textsc{SPIRA}\xspace}

\mdfdefinestyle{ResearchQuestion}{%
    linecolor=black,
    outerlinewidth=8pt,
    skipabove=\baselineskip,
    skipbelow=\baselineskip,
    innertopmargin=8pt,
    innerbottommargin=8pt,
    innerrightmargin=8pt,
    innerleftmargin=8pt,
}

\newcounter{RQ}
\renewcommand{\theRQ}{\arabic{RQ}}

\newcounter{SubRQ}[RQ] 
\renewcommand{\theSubRQ}{\theRQ.\arabic{SubRQ}}

\newenvironment{researchquestion}{%
  \refstepcounter{RQ}%
  \begin{mdframed}[style=ResearchQuestion]\label{rq:\theRQ}%
  \textbf{RQ\theRQ:}%
}{%
  \end{mdframed}%
}

\mdfdefinestyle{RevisitResearchQuestion}{%
    linecolor=black,
    outerlinewidth=8pt,
    skipabove=\parskip,
    skipbelow=\parskip,
    innertopmargin=8pt,
    innerbottommargin=8pt,
    innerrightmargin=8pt,
    innerleftmargin=8pt,
}

\crefformat{RQ}{#2RQ#1#3}
\Crefformat{RQ}{#2RQ#1#3}
\crefformat{SubRQ}{#2RQ#1#3}
\Crefformat{SubRQ}{#2RQ#1#3}

\mdfdefinestyle{ExpectedResult}{%
    linecolor=black,
    outerlinewidth=8pt,
    skipabove=\baselineskip,
    skipbelow=\baselineskip,
    innertopmargin=8pt,
    innerbottommargin=8pt,
    innerrightmargin=8pt,
    innerleftmargin=8pt,
}

\newcounter{ER}
\renewcommand{\theER}{\arabic{ER}}

\crefformat{ER}{#2ER#1#3}
\Crefformat{ER}{#2ER#1#3}

\newcounter{MethPhase}
\renewcommand{\theMethPhase}{\arabic{MethPhase}}

\newcounter{MethStep}[MethPhase] 
\renewcommand{\theMethStep}{\theMethPhase.\arabic{MethStep}}

\newcommand{\MethodologyPhase}{\refstepcounter{MethPhase}\label{phase:\theMethPhase}}

\newcommand{\MethodologyStep}{\refstepcounter{MethStep}\label{step:\theMethStep}}

\crefname{MethPhase}{phase}{phases}
\Crefname{MethPhase}{Phase}{Phases}
\crefname{MethStep}{step}{steps}
\Crefname{MethStep}{Step}{Steps}

\definecolor{drawio-blue}{HTML}{1ba1e2} %
\definecolor{drawio-gray}{HTML}{647687} %
\definecolor{drawio-green}{HTML}{6d8764} %
\definecolor{drawio-orange}{HTML}{fa6800} %
\definecolor{drawio-pink}{HTML}{99004D} %
\definecolor{drawio-purple}{HTML}{76608a} %
\definecolor{drawio-red}{HTML}{e51400} %
\definecolor{drawio-white}{HTML}{f9f7ed} %
\definecolor{drawio-black}{HTML}{36393d} %
\definecolor{drawio-yellow}{HTML}{e3c800} %

\definecolor{drawio-violet}{HTML}{6a00ff} %
\definecolor{drawio-magenta}{HTML}{dd0073} %
\definecolor{drawio-moss}{HTML}{008a00} %

\newcommand{\TikzSkewedSquare}[1][black]{%
  \tikz[baseline] {\draw[transform shape, black, fill={#1}] (0,0) rectangle (1.5ex,1.5ex);}%
}
\newcommand{\TikzSkewedCircle}[1][black]{%
  \tikz[baseline] {\draw[yshift=0.75ex, anchor=south, transform shape, black, fill={#1}] (0,0) circle (0.75ex);}%
}

\newcommand{\LegendColoredSquare}[3]{%
  \textcolor{#1}{#3}~\TikzSkewedSquare[#2]%
}
\newcommand{\LegendColoredCircle}[3]{%
  \textcolor{#1}{#3}~\TikzSkewedCircle[#2]%
}

\newcommand{\LegendColoredComponent}[2]{%
  \LegendColoredSquare{drawio-#1}{drawio-#1}{#2}%
}
\newcommand{\LegendBWComponent}[1]{%
  \LegendColoredSquare{drawio-black}{drawio-white}{#1}%
}
\newcommand{\LegendColoredLabel}[2]{%
  \LegendColoredCircle{drawio-#1}{drawio-#1}{#2}%
}

\mdfdefinestyle{Supervision}{%
    linecolor=black,
    outerlinewidth=8pt,
    innertopmargin=12pt,
    innerbottommargin=12pt,
    innerrightmargin=12pt,
    innerleftmargin=12pt,
}

\newcommand{\Quote}[1]{\textit{``#1''}}

\begin{document}


\title{
A Tale of Two Systems:\break
Characterizing Architectural Complexity on\break
Machine Learning-Enabled Systems
}
\titlerunning{Characterizing Architectural Complexity on ML-Enabled Systems}


\author{
Renato Cordeiro Ferreira\inst{1,2}\orcidID{0000-0001-7296-7091}
}

\authorrunning{Renato Cordeiro Ferreira}


\institute{%
Institute of Mathematics and Statistics (IME),
University of São Paulo (USP),
Brazil \\
\email{renatocf@ime.usp.br} \\
Jheronimus Academy of Data Science (JADS),
Tilburg University~(TiU) and Eindhoven University of Technology~(TUe),
The Netherlands \\
\email{r.cordeiro.ferreira@jads.nl}
}


\maketitle 


\begin{abstract}

How can the complexity of ML-enabled systems be managed effectively?
The goal of this research is to investigate how complexity affects
ML-Enabled Systems (MLES). To address this question, this research
aims to introduce a metrics-based architectural model to characterize
the complexity of MLES. The goal is to support architectural decisions,
providing a guideline for the inception and growth of these systems.
This paper brings, side-by-side, the architecture representation of
two systems that can be used as case studies for creating the
metrics-based architectural model: the \SPIRA and the \OG MLES.

\keywords{%
Software Metrics
\and Software Complexity
\and Machine Learning Enabled Systems
\and Machine Learning Engineering
\and MLOps
}%

\end{abstract}


\section{Introduction}
\label{sec:introduction}

\emph{Complexity} has been a subject of discussion since the early
days of the Software Engineering field~\cite{Brooks1975TheMan-Month}.
In the book ``Mythical Man Month'', Brooks introduces the concept of
\emph{essential} and \emph{accidental} complexity for software%
~\cite{Brooks1975TheMan-Month}:
the \emph{essential} exists intrinsically to satisfy the requirements
of the problem solved, whereas the \emph{accidental} may originate
from any external factors that influence the solution chosen.
Under this definition, an ML-enabled system (MLES) has high essential
complexity: it requires extra components in its architecture to
support data processing and model handling%
~\cite{Amershi2019SoftwareStudy,Benton2020MachineApplications}.
As a consequence, this also creates more opportunities for
introducing accidental complexity into the system.


While there are many recent surveys summarizing important gaps in the SE4AI
literature, only Giray \etal~highlights \emph{handling complexity} as an
open challenge for MLE~\cite{Giray2021AChallenges}.
\mbox{In an interview} study with 18 professional ML engineers%
~\cite{Shankar2022OperationalizingStudy},
Shankar et al. paper describes how participants \Quote{expressed an aversion
to complexity}, implying that practitioners consider it an issue.

This proposal presents a plan to create a \emph{metrics-oriented
architectural model to characterize the complexity of MLES}.
By using metrics to identify where complexity emerges in the software
architecture of MLES, this research aims to provide a method to avoid
pitfalls that make them fail to reach production.

\section{Research Questions}
\label{sec:research_questions}

To achieve the goal of this research, this PhD proposes
two main research questions. They follow the SMART principle%
~\cite{Verschuren2010DesigningDesign}, i.e., they should be
\emph{Specific, Measurable, Achievable, Relevant, and Time-Bound}.

\begin{researchquestion}
What are the measurable dimensions of complexity in
the architecture of MLES?
\end{researchquestion}

\Cref{rq:1} aims to explore metrics related to complexity available in
the literature. \emph{Software Metrics} have been thoroughly researched
by Software Engineering~\cite{Fenton2014SoftwareEdition}.
However, MLES also have their data and model dimensions%
~\cite{Sato2019ContinuousLearning}, which affect the software architecture.
Finding metrics that measure the data- and model-related complexity 
\emph{beyond code} is the biggest potential challenge for
answering~\cref{rq:1}.

\begin{researchquestion}
How can complexity metrics be operationalized over
the architecture of MLES?
\end{researchquestion}

\Cref{rq:2} aims to define a process to collect the metrics found
in \cref{rq:1}.
Metrics may be related to different abstraction levels of a system:
from code snippets, to components, to services%
~\cite{Fenton2014SoftwareEdition}.
Moreover, they can have varying levels of different quality attributes%
~\cite{Latva-Koivisto2001FindingModels},
such as validity, reliability, computability, intuitiveness,
ease of implementation, and independence of other metrics.
As a consequence, only some metrics found in \cref{rq:1} may
be practical or useful to collect. In particular, if two metrics
provide similar information, it will be preferable to use the best
according to the quality attributes. 
Therefore, the challenge for answering~\cref{rq:2} is twofold: creating a
dataset of representative MLES to collect the metrics, and then choosing which
subset of metrics to collect.

Measuring a metric may require different levels of access to the system
(e.g., having the codebase available for processing) or rely on different
representations of the system (e.g., creating a graph describing its data
flow). As a consequence, only some metrics found in \cref{rq:1} may be
practical to collect. In particular, if two metrics provide similar
information, it may be preferable to use the simpler.

\section{Methodology}\label{sec:research_methodology}

\begin{figure}[ht!]
\centering
\includegraphics[width=0.9\linewidth]{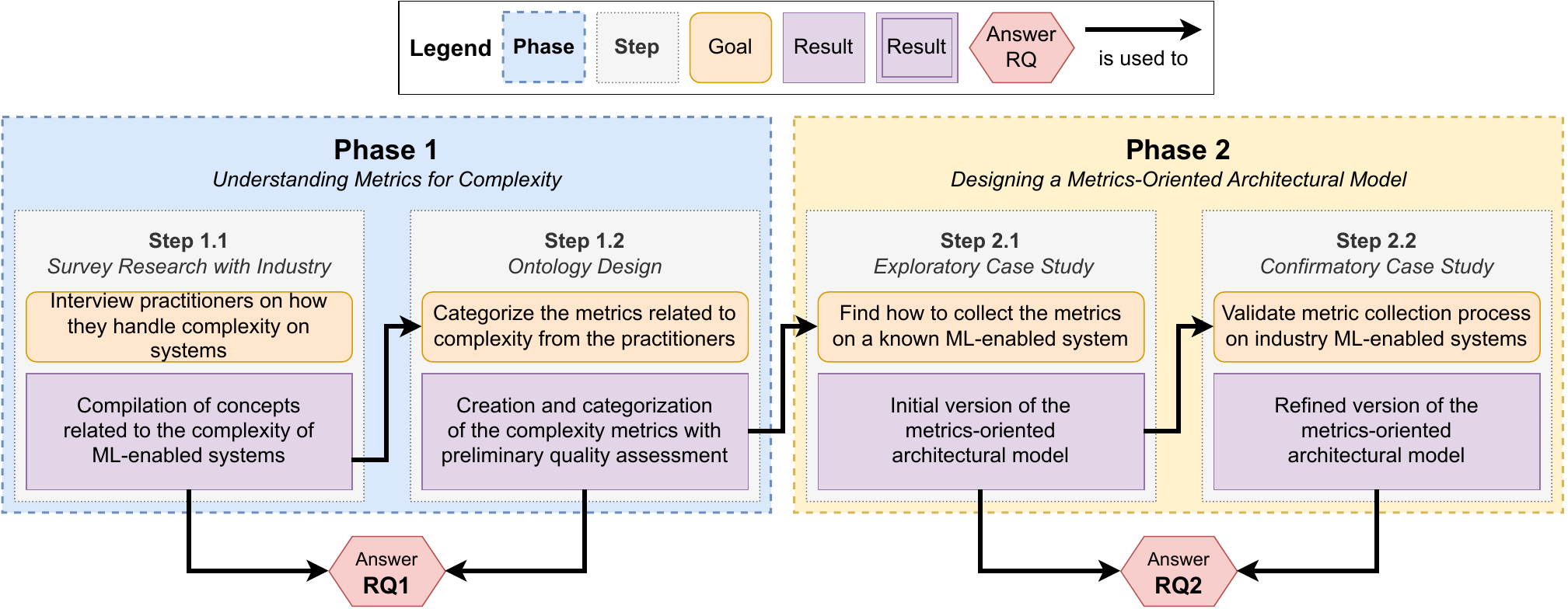}
\caption[%
  Research Methodology%
]{%
  \emph{Research Methodology}.
  The methodology is divided into two phases, each addressing a
  research question from \cref{sec:research_questions}.
  This methodology categorizes as empirical software engineering
  with a mixed-method approach.
}
\vspace{-0.25cm}
\label{fig:research_methodology}
\end{figure}

\Cref{fig:research_methodology} summarizes four steps, grouped into two
phases, of the methodology proposed for this research.

\subsection{Understanding Metrics for Complexity}
\label{subsec:methodology_phase_1}
\MethodologyPhase

To answer \cref{rq:1}, this research will start studying
\emph{the complexity of software systems} and \emph{metrics}.
For that, \cref{phase:1} has been divided into two steps:
(1) to understand how practitioners address complexity in their systems; and
\MethodologyStep
(2) to describe and categorize related metrics based on the literature.
\MethodologyStep

\subsubsection{Survey Research with Industry.}
Survey research proposes selecting a representative sample from
a well-defined population, identifying characteristics from it
(via questionnaires or structured interviews), and then applying
data analysis techniques to generalize the results%
~\cite{Easterbrook2008SelectingResearch}.
This method helps to answer a question about a particular target
population, such as professional developers working on software
companies, without polling every member of the population.

A survey allows a practical way to map how practitioners address
complexity while creating MLES, as is the goal of \cref{step:1.1}.
This approach is \emph{inductive} rather than \emph{deductive}, i.e.,
first it proposes to understand what practitioners consider while
designing their systems, then it tries to create a theory around it.
Since MLES have their data and model dimensions%
~\cite{Sato2019ContinuousLearning}, these characteristics will likely
reflect during the characterization of the complexity of MLES.



\subsubsection{Ontology Design.}
Making an ontology results in a well-defined categorization of the metrics%
~\cite{NoyOntologyOntology},
as it is the goal of \cref{step:1.2}. The ontology will focus on describing
different \emph{types} of metrics~\cite{Fenton2014SoftwareEdition},
as well as possible \emph{quality attributes} associated with them%
~\cite{Latva-Koivisto2001FindingModels}. Some properties will be
related to the \emph{applicability} of the metrics, which will be explored
only on~\cref{phase:2}. Therefore, the knowledge base will be fully complete
only by the end of the research.

An ontology can be created via \emph{ontology design}%
~\cite{NoyOntologyOntology}. An ontology
is the structure behind a knowledge base, including \emph{classes},
\emph{properties}, and \emph{restrictions}~\cite{NoyOntologyOntology}.
The ontology will focus on describing different \emph{types} of
metrics~\cite{Fenton2014SoftwareEdition}, as well as possible
\emph{quality attributes} associated with them%
~\cite{Latva-Koivisto2001FindingModels}. 


\subsection{Designing a Metrics-Oriented Architectural Model}
\label{subsec:methodology_phase_2}
\MethodologyPhase

To answer \cref{rq:2}, this research will focus on selecting a subset of the
metrics from \cref{phase:1} to compose the \emph{metrics-oriented architectural
model}. For that, \cref{phase:2} has been divided into two steps:
(1) to find how to collect the metrics on a known MLES; and
\MethodologyStep
(2) to validate the metric collection process on industry MLES.
\MethodologyStep

\subsubsection{Exploratory Case Study.}
Making an exploratory case study provides a framework to propose an answer
to~\cref{rq:2}~\cite{Easterbrook2008SelectingResearch}.
The exploratory case study will have the following study proposition:
\emph{``What is a process to operationalize the collection of complexity
metrics on an MLES?''}, as it is the goal of \cref{step:2.1}.
To achieve that, it will rely on two \emph{typical cases}%
~\cite{Easterbrook2008SelectingResearch}:
the \SPIRA MLES~\cite{Ferreira2022SPIRA:Detection,Lawand2025IsDomain}
and the \OG MLES~\cite{Ferreira2025MLOpsDomain}, which are currently
being developed by the author alongside his PhD research.

The familiarity with the codebase should allow the author to focus
on operationalizing the metric collection rather than learning how the
project works. The goal is twofold:
document \emph{the process of metric collection},
to achieve the study proposition; and
determine \emph{the quality attributes of the metrics},
to refine the ontology made on \cref{step:1.2}.

By choosing which metrics are worth collecting, and documenting
the requirements to collect them, the result will be the initial version
of the \emph{metrics-oriented architectural model}.

\subsubsection{Confirmatory Case Study.}
Making a confirmatory case study provides a framework to validate the proposed
answer to~\cref{rq:2}~\cite{Easterbrook2008SelectingResearch}.
The confirmatory case study will repeat the same study proposition:
\emph{``What is a process to operationalize the collection of complexity
metrics on an MLES?''}, as it is the goal of \cref{step:2.2}.
To achieve that, this research can rely on databases of open source
repositories compiled by previous works from the literature, such as
the 262 open-source ML products mined by Nahar \etal~from GitHub%
~\cite{Nahar2025TheProducts}.

Production MLES can have various architectures, with different combinations
of components. The diversity of codebases should allow the researchers to
validate the robustness of the metric collection process. Two types of
data will be revisited: \emph{the process of metric collection}, to achieve
the study proposition; and \emph{the quality attributes of the metrics},
to refine the ontology made on \cref{step:1.2}.

By reviewing which metrics are worth collecting and how to collect them,
the result will be a refined version of the
\emph{metrics-oriented architectural model}.

\section{Preliminary Results}
\label{sec:the_spira_system}

As described in \cref{sec:research_methodology}, this research proposes to 
use two case studies as typical cases to develop the first version of the
\emph{metrics-oriented architectural model}: the \SPIRA MLES and the \OG MLES.
To be able to compare the complexity of both systems, the first step was to 
describe them under a similar representation. 

\begin{figure}[p]
  \centering
  \includegraphics[width=\linewidth]{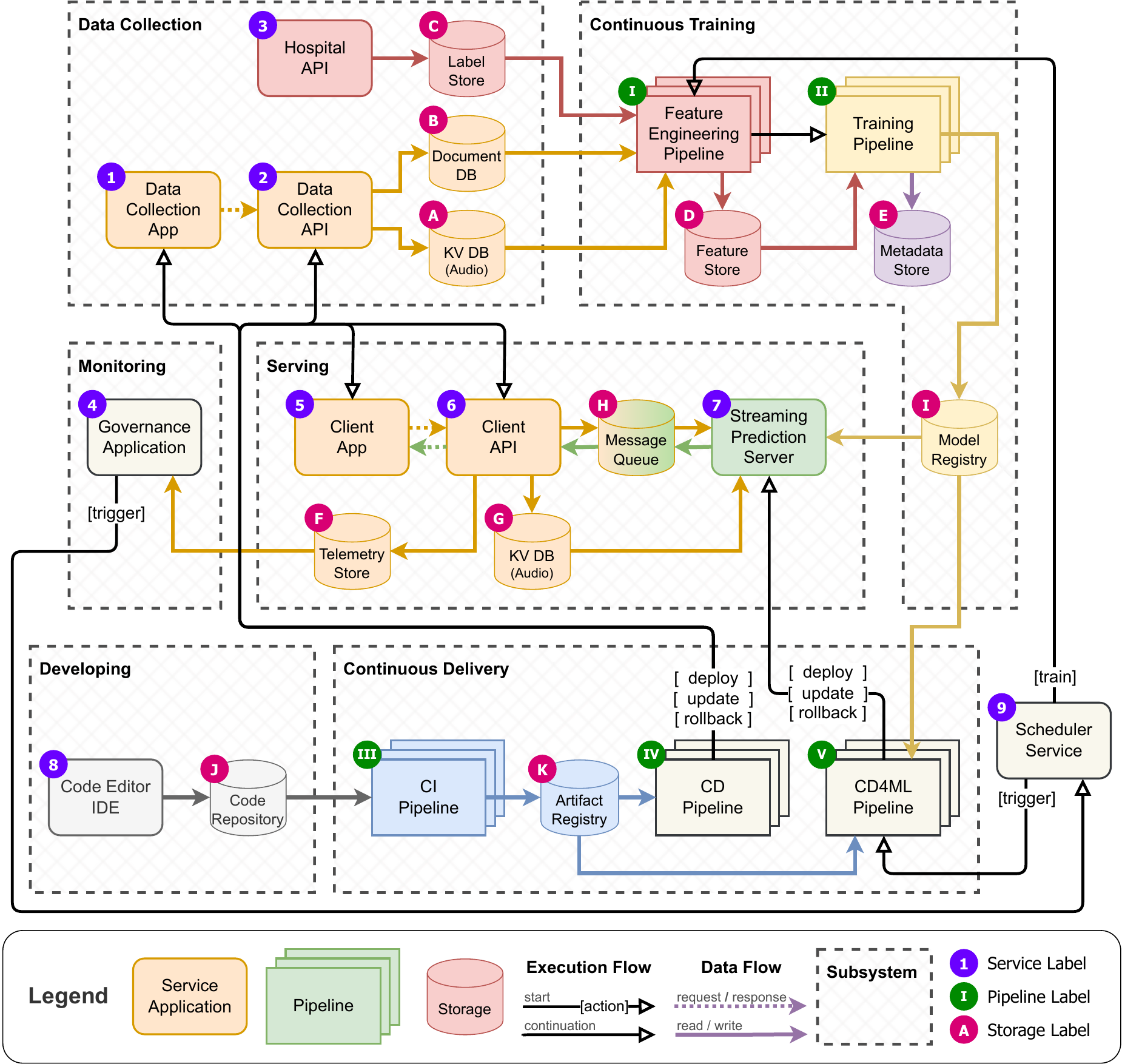}
  \caption[%
    Architecture of the \SPIRA ML-Enabled System%
  ]{%
    \textbf{System Architecture of \SPIRA.}
    This architecture follows the notation presented at the PhD Symposium
    at \href{https://conf.researchr.org/home/cain-2025}{CAIN 2025}%
    ~\cite{Ferreira2025ASystems}, while the \SPIRA architecture was
    submitted to \href{https://sadis2025.smartarch.cz/}{SADIS 2025}%
    ~\cite{Lawand2025IsDomain}.
    Rectangles represent \textbf{applications} or \textbf{services},
      which execute continuously.
    Stacked rectangles represent \textbf{pipelines},
      which execute a task on demand.
    Lastly, cylinders represent \textbf{data storage},
      which may be databases of any type.
    Components are connected by arrows.
    Black arrows with a hollow tip illustrate the \textbf{execution flow}.
      They start and end in a component.
      Labeled arrows represent the trigger that starts a workflow,
      whereas unlabeled arrows represent the continuation of an
      existing workflow.
    Colored arrows with a filled tip illustrate the \textbf{data flow}.
    They appear in two types:
      solid arrows going to and from a data storage represent
      write and read operations, respectively;
      dotted arrows represent a sync or async request-response
      communication between components.
    Components are colored according to the data they produce:
      \mbox{\LegendColoredComponent{orange}{raw data}},
      \mbox{\LegendColoredComponent{red}{ML-specific data}},
      \mbox{\LegendColoredComponent{gray}{source code}},
      \mbox{\LegendColoredComponent{blue}{executable artifacts}},
      \mbox{\LegendColoredComponent{yellow}{ML models}},
      \mbox{\LegendColoredComponent{purple}{ML training metadata}},
      \mbox{\LegendColoredComponent{green}{ML model predictions}}, and
      \mbox{\LegendColoredComponent{pink}{ML model metrics}}.
      Remaining \LegendBWComponent{standalone components} orchestrate
      the execution of others.
    Components are also grouped into \textbf{subsystems}.
    \LegendColoredLabel{violet}{Numbers},
    \LegendColoredLabel{moss}{Roman numerals}, and
    \LegendColoredLabel{magenta}{letters} are used as labels.
  }
  \label{fig:spira_architecture}
\end{figure}

\SPIRA is a MLES for prediagnostic of insufficiency respiratory
via speech analysis. It is being developed since 2020 as a
\href{https://bv.fapesp.br/en/auxilios/113158/}{multidisciplinary project}
under the University of São Paulo (USP), in Brazil. The project aims to create
a tool to assist physicians and nurses  to find early signs of respiratory
insufficiency, thus helping physicians to identify which patients need
urgent treatment. \Cref{fig:spira_architecture} describes the architecture
of \SPIRA, as submitted for \href{https://sadis2025.smartarch.cz/}{SADIS 2025}%
~\cite{Lawand2025IsDomain}.

\begin{figure}[p]
  \centering
  \includegraphics[width=0.96\linewidth]{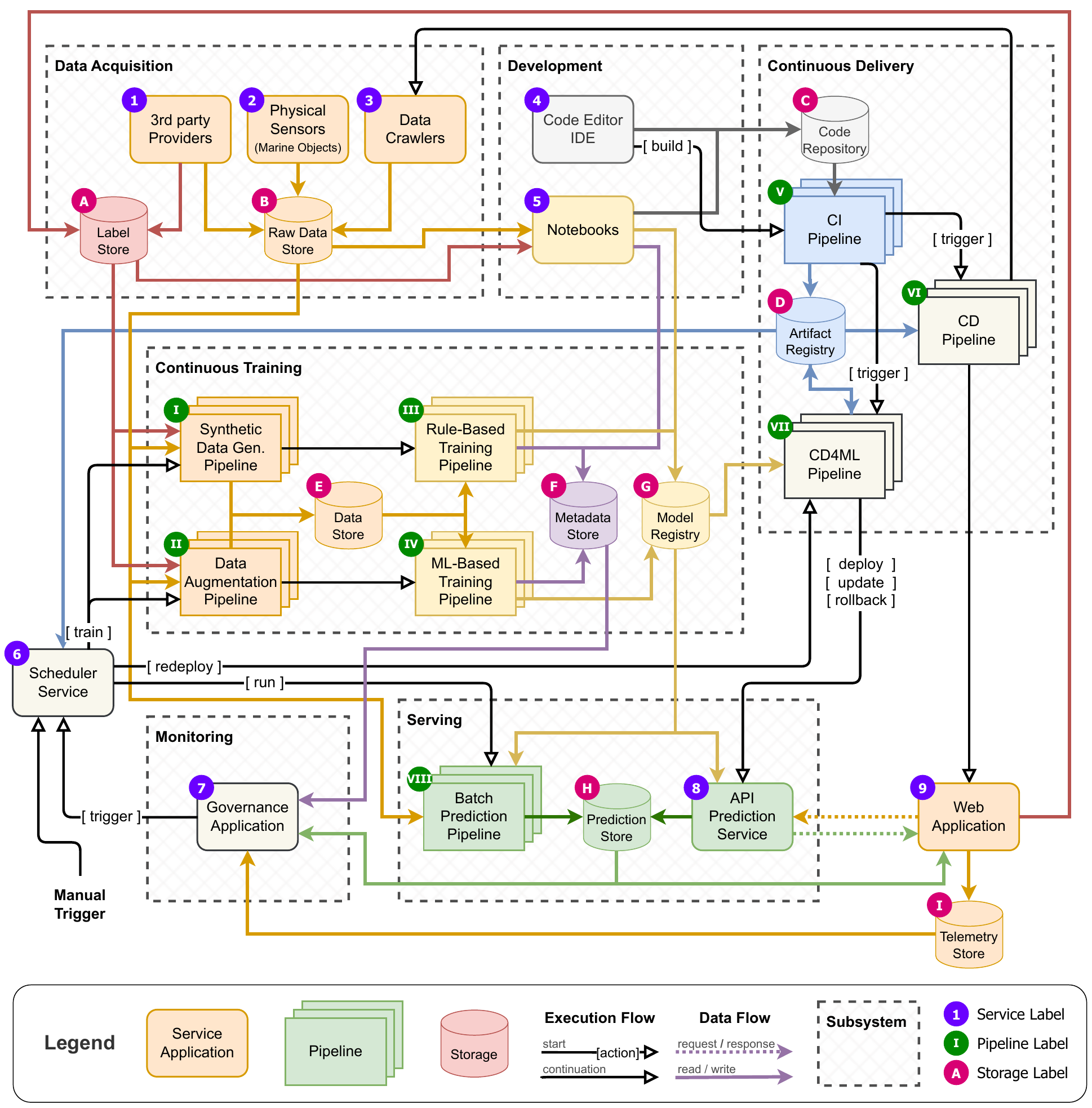}
  \caption[
    Architecture of the \OG ML-Enabled System%
  ]{
    \textbf{System Architecture of \OG.}
    This architecture follows the notation presented at the PhD Symposium
    at \href{https://conf.researchr.org/home/cain-2025}{CAIN 2025}%
    ~\cite{Ferreira2025ASystems}, while the \OG architecture was submitted
    to \href{https://www.summersoc.eu/}{SummerSOC 2025}%
    ~\cite{Ferreira2025MLOpsDomain}.
    Rectangles represent \textbf{applications} or \textbf{services},
      which execute continuously.
    Stacked rectangles represent \textbf{pipelines},
      which execute a task on demand.
    Lastly, cylinders represent \textbf{data storage},
      which may be databases of any type.
    Components are connected by arrows.
    Black arrows with a hollow tip illustrate the \textbf{execution flow}.
      They start and end in a component.
      Labeled arrows represent the trigger that starts a workflow,
      whereas unlabeled arrows represent the continuation of an
      existing workflow.
    Colored arrows with a filled tip illustrate the \textbf{data flow}.
    They appear in two types:
      solid arrows going to and from a data storage represent
      write and read operations, respectively;
      dotted arrows represent a sync or async request-response
      communication between components.
    Components are colored according to the data they produce:
      \mbox{\LegendColoredComponent{orange}{raw data}},
      \mbox{\LegendColoredComponent{gray}{source code}},
      \mbox{\LegendColoredComponent{blue}{executable artifacts}},
      \mbox{\LegendColoredComponent{red}{ML-specific data}},
      \mbox{\LegendColoredComponent{yellow}{ML models}},
      \mbox{\LegendColoredComponent{purple}{ML training metadata}},
      \mbox{\LegendColoredComponent{green}{ML model predictions}}, and
      \mbox{\LegendColoredComponent{pink}{ML model metrics}}.
      Remaining \LegendBWComponent{standalone components} orchestrate
      the execution of others.
    Components are also grouped into \textbf{subsystems}.
    \LegendColoredLabel{violet}{Numbers},
    \LegendColoredLabel{moss}{Roman numerals}, and
    \LegendColoredLabel{magenta}{letters} are used as labels.
  }
  \label{fig:og_architecture}
\end{figure}

\OG is a MLES for anomaly detection in the maritime domain. It is being
developed since 2023 as part of the \href{https://marit-d.eu/}{MARIT-D}
European project under the Jheronimus Academy of Data Science (JADS),
in the Netherlands. The project aims to create a tool to enhance the
efficiency and effectiveness of law enforcement agencies (LEAs)
across the European Union. \Cref{fig:og_architecture} describes the
architecture of \OG, as submitted for \href{https://www.summersoc.eu/}{
SummerSOC 2025}~\cite{Ferreira2025MLOpsDomain}.

\Cref{fig:spira_architecture,fig:og_architecture} use the architecture
representation published by the author in the Doctoral Symposium of
\href{https://conf.researchr.org/home/cain-2025}{CAIN 2025}%
~\cite{Ferreira2025ASystems}, which is an extension of the
reference architecture by Kumara \etal, available since 2023%
~\cite{Kumara2023RequirementsIndustry} and accepted for publication at
\href{https://conf.researchr.org/home/ecsa-2025}{ECSA 2025}%
~\cite{Kumara2025MLOpsIndustry}.

\section{Challenges and Open Questions}
\label{sec:challenges_open_questions}

This PhD started at University of São Paulo (USP) in 2020, amidst
the COVID-19 pandemic. The author was employed in a full-time job as
\emph{Principal ML Engineer}. This expertise led him to participate in
the \SPIRA project, which led him to supervise many students to develop
the MLES described in \cref{fig:spira_architecture}.

This PhD started a new phase at the Jheronimus Academy of Data Science
(JADS) in 2023, after an opportunity of a double degree with the Netherlands.
The author became employed as a full-time \emph{Scientific Programmer} for
the \mbox{MARIT-D} project. This opportunity led him to develop the
\OG MLES described in \cref{fig:og_architecture}.

The development of \SPIRA and \OG is intertwined with this PhD research, and
both projects can be excellent case studies for \cref{step:2.1}. Currently,
\cref{step:1.1} is being executed with the help of two master students, whose
research includes a broader interview-based qualitative study about MLOps.
Therefore, the research is converging to a point where it will be possible
to measure the first metrics of complexity about the systems, thus creating
the first version of the \emph{metrics-oriented architectural model}.

Given this context, some open questions for the research are:
\begin{itemize}
  \item How to take advantage of the familiarity with both systems to
        explore the dimensions of complexity of MLES?
  \item Is it acceptable to choose metrics for the first version of the
        \emph{metrics-oriented architectural model}
        if they can be measured in these systems?
  \item How to mitigate experimental bias and other threats to validity 
        when creating the \emph{metrics-oriented architectural model},
        considering the familiarity with the case studies?
\end{itemize}


%
%
%
\bibliographystyle{splncs04}
\bibliography{mendeley}

\end{document}